\documentclass[mathleft
]{an}
\usepackage{graphicx}
\usepackage{times}
\newcommand{\kms}{km\,s$^{-1}$}

\begin{document}

\Pagespan{789}{}
\Yearpublication{2013}%
\Yearsubmission{2012}%
\Month{11}%
\Volume{334}%
\Issue{5}%

\title{Long-term photometry of three active red giants in \\close binary systems:
   V2253~Oph, IT~Com and IS~Vir}

\author{K.~Ol\'ah\inst{1}\thanks{Corresponding author:
  \email{olah@konkoly.hu}}, A.~Mo\'or\inst{1}, K.~G.~Strassmeier\inst{2}, T.~Borkovits\inst{1,3,4}, 
\and T.~Granzer\inst{2}}

\titlerunning{Long-term photometry of three active red giants in close binary systems}
\authorrunning{Ol\'ah et al.}

\institute{
Konkoly Observatory MTA CsFK, Konkoly Thege M. u. 15/17, H-1121 Budapest, Hungary
         \and
Leibniz Institute for Astrophysics Potsdam (AIP), An der Sternwarte 16, D-14482 Potsdam, Germany
	  \and
Baja Astronomical Observatory,  Szegedi \'ut Kt.~766, H-6500 Baja, Hungary
	  \and
 ELTE Gothard-Lend\"ulet Research Group, H-9700 Szombathely, Hungary}

\received{2012}
\accepted{2012}
\publonline{later}


\abstract{We present and analyze long-term optical photometric measurements of the three active stars V2253\,Oph, IT\,Com and IS\,Vir. All three systems are single-lined spectroscopic binaries with an early K giant as primary component but in different stages of orbital-rotational synchronization. Our photometry is supplemented by 2MASS and WISE near-IR and mid-IR magnitudes and then used to obtain more accurate effective temperatures and extinctions. For V2253\,Oph and IT\,Com, we found their spectral energy distributions consistent with pure photospheric emission. For IS\,Vir, we detect a marginal mid-IR excess which hints towards a dust disk. The orbital and rotational planes of IT\,Com appear to be coplanar, contrary to previous findings in the literature. We apply a  multiple frequency analysis technique to determine photometric periods, and possibly changes of periods, ranging from days to decades. New rotational periods of 21.55$\pm$0.03\,d, 65.1$\pm$0.3\,d, and 23.50$\pm$0.04\,d were determined for V2253\,Oph, IT\,Com, and IS\,Vir, respectively. Splitting of these periods led to tentative detections of differential surface rotations of $\delta P/P\approx0.02$ for V2253\,Oph and $0.07$ for IT\,Com. Using a time-frequency technique based on short-term Fourier transforms we present evidence of cyclic light variations of length $\approx$10\,yrs for V2253\,Oph and 5-6\,yrs for IS\,Vir. A single flip-flop event has been observed for IT\,Com of duration 2--3\,yrs.  Its exchange of the dominant active longitude had happened close to a time of periastron passage, suggesting some response of the magnetic activity from the orbital dynamics. The 21.55-d rotational modulation of V2253\,Oph showed phase coherence also with the orbital period, which is 15 times longer than the rotational period, thus also indicating a tidal feedback with the stellar magnetic activity.
}

\maketitle

\section{Introduction}\label{S1}

During the last two decades numerous small and specialized surveys were carried out in order to identify new cool stars across the H-R diagram that show photospheric and/or chromospheric signs of magnetic activity (e.g. Lockwood et al. \cite{lock}, Baliunas et al. ~\cite{bal}, Henry et al. \cite{henry96}, Strassmeier et al. ~\cite{kpno}, Henry et al.~\cite{henry00}, De~Medeiros et al.~\cite{dem2002}, Wright et al.~\cite{ccps}, White et al.~\cite{white2007}, Strassmeier et al. \cite{binaries}). The next and still ongoing step was to obtain more accurate basic astrophysical parameters for these stars and to conduct time-resolved photometric and spectroscopic monitoring in order to determine rotational periods and orbital elements.  The seemingly most interesting objects were followed up with dedicated campaigns with automatic telescopes at even higher cadence in order to sample their starspot evolution  (e.g. Ol\'ah et al.~\cite{HKLac}, J\"arvinen et al.~\cite{V889}, Berdyugina \& Henry \cite{HR1099}, Strassmeier et al.~\cite{bepsc}, Savanov \& Strassmeier \cite{V1355}, Korhonen et al.~\cite{3stars}, Lehtinen et al.~\cite{lqhya}).  

A valuable tool to extend the time-base of the datasets for studying the long-term behavior of magnetic activity is the use of plate archives spanning decades or even a century. As examples we mention that Phillips \& Hartmann (\cite{phillips_hartmann}) determined first time long-term starspot activity of BY~Dra and CC~Eri from Harvard plates from half a century between 1900-1950, and Fr\"ohlich et al. (\cite{frohlich2, frohlich1}) used the Sonneberg plate collection for studying the long-term behaviour of the solar-type single active star EK~Dra and the K-giant primary of the close binary HK~Lac, respectively. Vogt et al. (\cite{nvogt}) found several new variables on the Sonneberg plate collection of 34 years, most of them are early type stars from late-B to A, showing slow long-term changes on the timescale of a few thousand days. Recently Balona (\cite{balona}) found rotational modulation on A stars from {\it Kepler} data which might originate from starspots as one possibility, and if this would be verified than longer-term magnetic variability could also be behind the slow changes shown by Vogt et al.'s (\cite{nvogt}) early type new variables. The DASCH (Digital Access to a Sky Century at Harvard) project recently resulted in the discovery of three K-giants' long-term variability by Tang et al (\cite{tang}), which are very probably active stars, similar to those studied in this paper.

Our main aim is to find evidence for a relation of photospheric starspot activity with the evolutionary state of the star or the orbital properties of the binary system.  Does the rearrangement of the internal structure at the base of the red-giant branch reflect onto changes of the dynamo properties? Or, can tidal interaction dissipate orbital energy into magnetic energy? In this paper, we report on follow-up observations of three chromospherically active binaries, each containing a K-giant star.

\begin{figure*}[!tbh]
{\bf a} \hspace{55mm}{\bf b} \hspace{55mm}{\bf c} \\
\includegraphics[width=5.5cm]{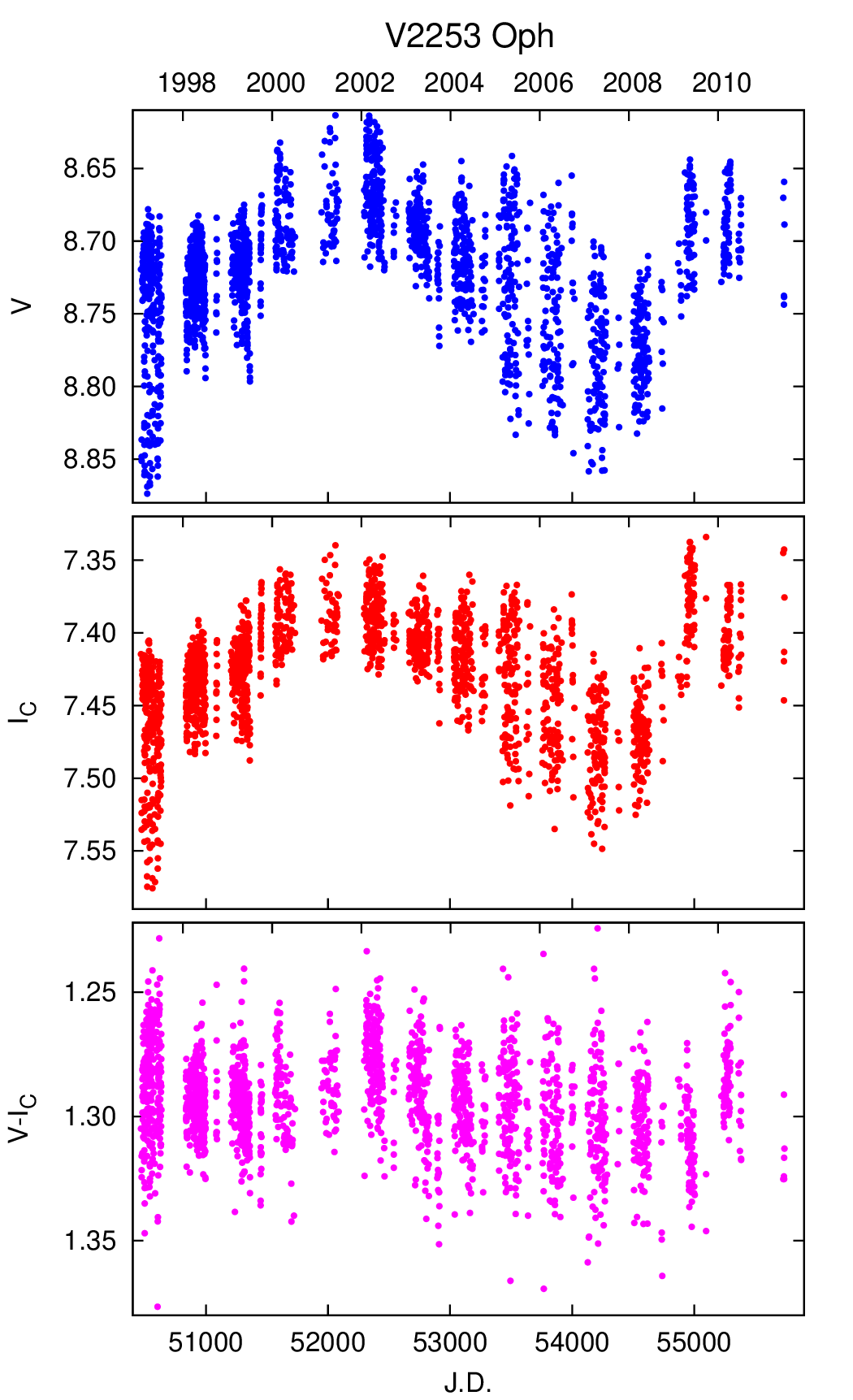}
\includegraphics[width=5.5cm]{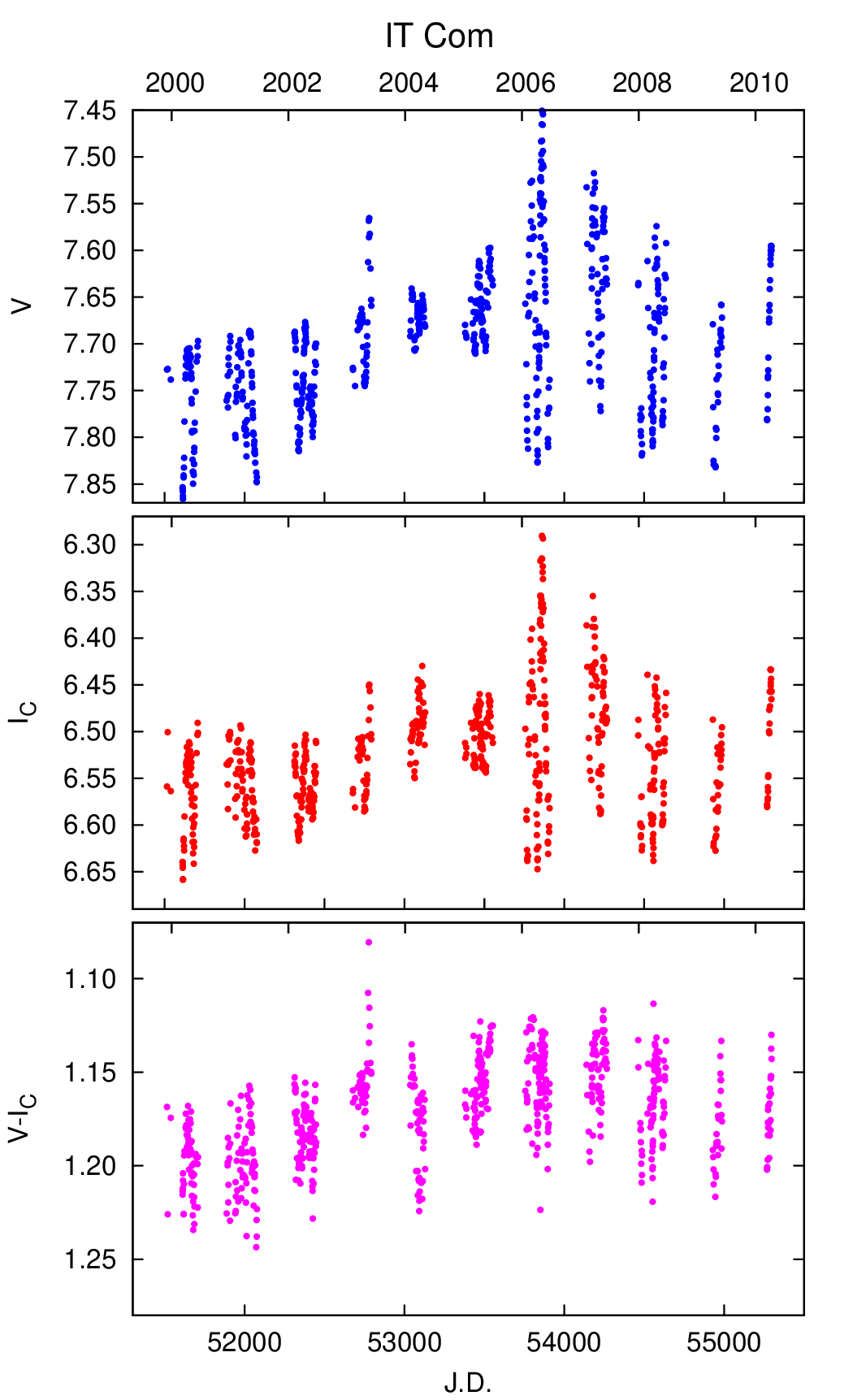}
\includegraphics[width=5.5cm]{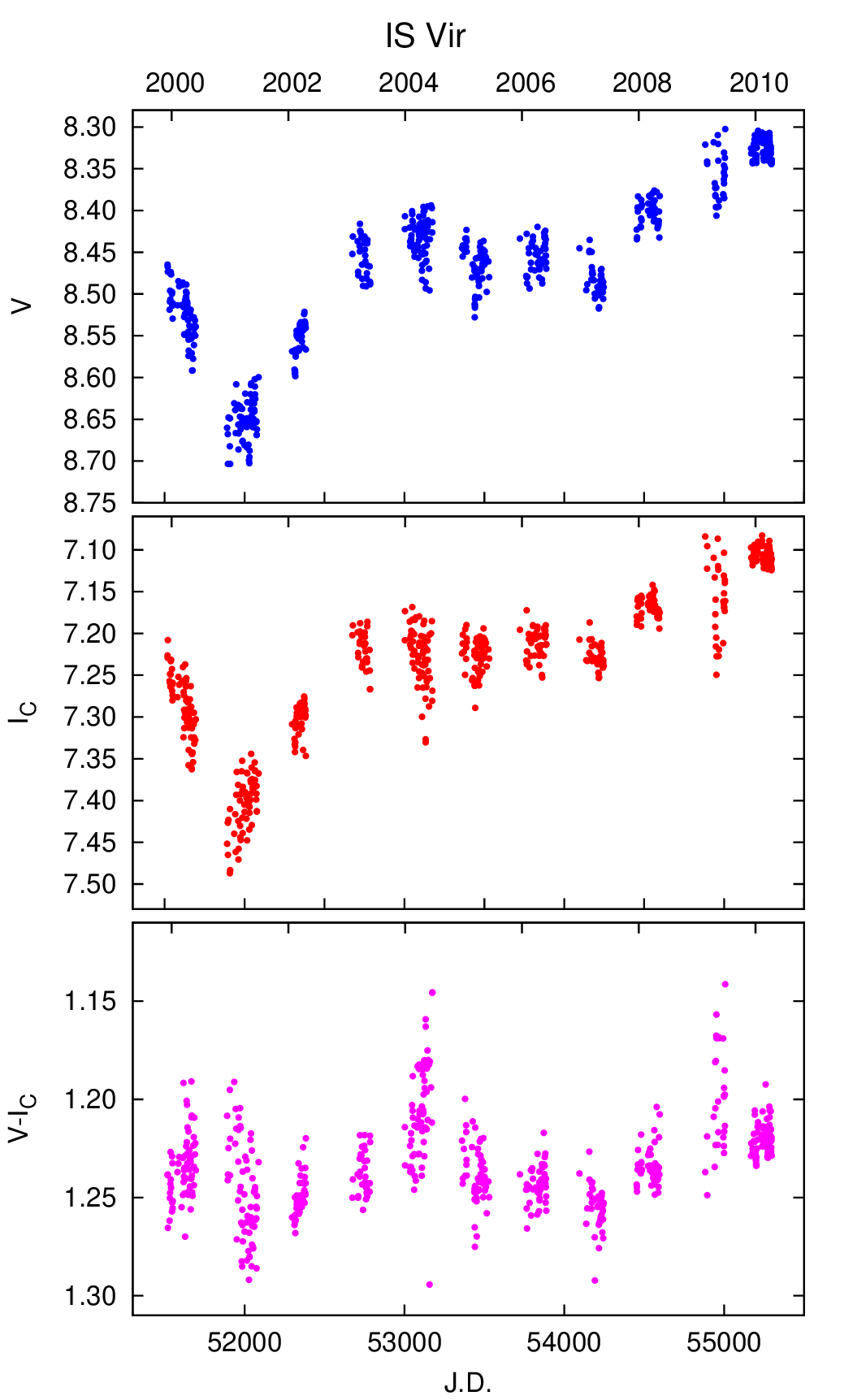}
\caption{Overview of the Vienna-Potsdam APT light curves of the three binaries in this paper; \emph{a.} V2253~Oph,  \emph{b.} IT~Com and \emph{c.} IS ~Vir. The APT took at least one $VI_C$ data point every clear night for the past 15 years. In the figure, target stars are arranged in columns.  $V$ magnitude is plotted in the top row, $I_C$ magnitudes in the middle row, and $V-I_C$ color in the bottom row. }
\label{lightcurves}
\end{figure*}

The spectral types, effective temperatures, and radii (when known) of the primary stars in these binaries are similar to each other. However, their degree of rotational synchronism to the orbital revolution is markedly different. V2253\,Oph is a short-period asynchronously-rotating star in a binary with an orbital period 15 times longer than the stellar rotation period. Its orbit is only slightly eccentric, if at all, because zero eccentricity is within 3$\sigma$ of the value given by Fekel et al. (\cite{fekel1}). Thus, the period difference could not be due to pseudosynchronization. IT\,Com is also an asynchronously rotating star but its rotation period is longer than the orbital period by about 10\%. Its inclination of the rotational axis is coplanar to the orbit, which itself is highly eccentric (Griffin \cite{griffin88}). The third system, IS~Vir, is a synchronized giant in a marginally eccentric orbit (Fekel et al. \cite{fekel1}).

In the following, we first summarize the relevant literature for the three binary systems in Sect.~\ref{S2}. In Sect.~\ref{S3}, we describe our new observations and present the tools that will later be applied in Sect.~\ref{S4} and Sect.~\ref{S5}.  Sect.~\ref{S6} summarizes and discusses our results.

\section{The program stars: V2253\,Oph, IT\,Com, and IS\,Vir}\label{S2}

Table~\ref{table:1} summarizes some of the relevant basic parameters for the three systems for quick reference.

The light variability of V2253\,Oph (HD\,152178) was discovered by Hooten \& Hall (\cite{hooten_hall}) who determined a rotational period of 22.35$\pm0.05$\,d from photometry. Their observations were initiated by the spectral classification of Houk (\cite{houk}) who noted Ca\,{\sc ii} H\&K emission and found the star to be a single-lined (SB1) spectroscopic binary. She also derived a spectral type of G8/K0  (but erroneously of luminosity class~V). The correct spectral type of the primary and the first orbital elements were published in the second CABS catalog (Strassmeier et al. \cite{CABSII}) showing a significantly longer orbital period of 314.5\,d. The most recent orbital elements are given in Fekel et al. (\cite{fekel2}) and basically confirmed the original orbit.  Strassmeier (\cite{Ca_sp}) and Strassmeier et al. (\cite{CaII}) determined absolute emission-line fluxes of the strong Ca\,{\sc ii} H\&K emissions and ROSAT detected the X-ray flux of the system (Dempsey et al. \cite{dem:lin}).

IT\,Com (HD\,118234) was first mentioned as a spectroscopic binary by Griffin (\cite{griffin88}) who published an SB1 orbital solution with a high eccentricity ($e$=0.589).  Together with unpublished photometry, he concluded that the primary should be an early K giant. Strassmeier (\cite{Ca_sp}) and Strassmeier et al. (\cite{CaII}) had this object on their target list and detected strong Ca\,{\sc ii} H\&K emission lines while ROSAT detected the system in X-rays (Dempsey et al. \cite{dem:lin}). The photometric variability  with a period of 64$\pm$1 days was discovered by Henry et al. (\cite{henry_etal}). This period implied that IT\,Com rotates slightly asynchronous with respect to the orbital period of $P_{\rm orb}$=59.054\,d. Henry et al. (\cite{henry_etal}) also noticed the strongly variable photometric amplitude between zero and 0\fm2. Glebocki \& Stawikowski (\cite {glebockistawikowski97}) stated that the orbit is not coplanar, $i_{\rm orb}=13\degr$ while $i_{\rm rot}=34\degr$; the authors found a general misalignment of the primaries of the asynchron systems to their orbits from observations and statistical analysis. However, the error of the inclination determinations are between 5-9 degrees for systems like IT\,Com, which may question some of the individual results.
Finally, IT\,Com was observed with the ELODIE spectrograph at OHP in the survey of Soubiran et al. (\cite{soubiran}), who derived atmospheric parameters, distance, and space velocities  together with a large sample of mostly clump giant stars.

\begin{table*}
\caption{The program stars.}\label{table:1}
\begin{tabular}{llll}
\hline\hline\noalign{\smallskip}
Parameter & V2253~Oph & IT~Com & IS~Vir \\
\hline\noalign{\smallskip}
  HD						 & 152178 & 118234 & 113816 \\
 $\langle V\rangle$ (mag)          & 8.5 & 7.4 & 8.3 \\
  M-K class                                 & K0III & K0III & K2III \\
  T$_\mathrm{eff}$ (K)               & 4900$\pm$100 & 4700 &  4700$\pm$77\\
  $\log g$                                    & (2.5) & 3.1 & 2.45$\pm$0.21 \\
  $v\sin i$ (\kms )                        & 28.8$\pm1$ & 5.3 & 5.9$\pm1$ \\
  $R\sin i$ (R$_\odot$)               & 12.7 & 7 & 2.7 \\
  $i$ ($^\circ$)                            & 63 & 13 & 13$\pm$4 \\
  $[$Fe/H$]$                               & \dots & $-0.19$& $-0.11\pm0.09$ \\
  Orbit                                         & SB1 & SB1 & SB1 \\
  $P_{\rm orb}$ (days)                & 314.47$\pm0.18$ & 59.054$\pm0.004$ &  23.655$\pm0.001$  \\
  $ e$                                          & 0.024$\pm0.009$ & 0.589$\pm0.010$  &  0.022$\pm0.008$ \\
  K (\kms)                                   & 14.32$\pm0.15$ & 9.66$\pm0.15$ & 6.598$\pm0.054$ \\
  $f(m)$                                      &  0.096$\pm$0.003 & 0.0029$\pm$0.0002 &  0.00070$\pm$0.00002 \\
  $a_1\sin i$ ($10^6$km)            &  61.92$\pm0.64$  &  6.34$\pm0.12$ &  2.146$\pm0.017$ \\
 $d$ (pc)                                     & 360$\pm105$ & 185$\pm25$ &  250$\pm50$\\
\noalign{\smallskip}\hline
\end{tabular}
\vspace{1mm}

\noindent {Source of orbital elements: Fekel et al. (\cite{fekel2}) for V2253\,Oph, Griffin (\cite{griffin88}) for IT\,Com and Fekel et al. (\cite{fekel1}) for IS\,Vir. The distance is from the revised \emph{Hipparcos} parallax (van Leeuven \cite{hipparcos}). The other parameters are for orientation only and were taken from the CABS-III catalog (Eker et al. \cite{CABSIII}) and references therein.}
\end{table*}

The third target, IS\,Vir (HD\,113816) is the best studied of the three targets in this paper. Fekel et al. (\cite{fekel1}) summarized the various measurements of this object from X-rays to optical photometry and spectroscopy that was published until 2002. They obtained a new, well-determined orbital solution with synchronous rotation and found a very low orbital and rotational inclination of this SB1 binary. From 12 years of photometry, cyclic light variability of the order of 7--8\,yrs was discovered. Detailed photospheric abundances of IS\,Vir were published by Katz et al. (\cite{katz}) and by Morel et al. (\cite{morel}). Both analysis found near-solar abundance with perhaps a slight overabundance of Na, Mg, Al and Ca with respect to Fe, by 0.1-0.2 dex; but the derived Fe/H ratio is quite different (0.04--0.09 and $-0.11\pm0.09$, respectively. 

\section{Observations and data-analysis methods}\label{S3}

\subsection{New APT observations}

We present altogether 3,191 new photometric $VI_C$ data points. All observations were taken with the 0.75m Vienna-Potsdam Automatic Photoelectric Telescope (APT) {\sl Amadeus} at Fairborn Observatory in southern Arizona (Strassmeier et al. \cite{apt0}). For V2253\,Oph, 1,883  measures were available from between June 1997 and June 2011. For IT\,Com, 698 measures from December 1999 through April 2010 and for IS\,Vir 610 measures from December 1999 through January 2011 could be used. These data are plotted in Fig.~\ref{lightcurves} for reference. Observations of V2253\,Oph from the first half of 1997 were published earlier (Strassmeier et al. \cite{apt}) but are added to the present dataset. An independent continuous 12-yr photometric dataset for IS\,Vir was published and analyzed by Fekel et al. (\cite{fekel1}) who started observing 10 years prior to our APT campaign with two seasons overlapping. We combine these two data sets and use it for parts of the present analysis. Our {\sl Amadeus} APT photometry had larger than usual scatter in the years between 2007 and 2009 due to target acquisition problems (see Strassmeier et al. \cite{binaries} for the details). However, the typical external precision of {\sl Amadeus} has been around 4--6~mmag depending on the brightness of the target. Photometry from literature sources are sparse for the other two systems, V2253\,Oph and IT\,Com, and are not taken into account because we would scrutinize the homogeneity of the APT data. Note that the IS\,Vir data of Fekel et al. (\cite{fekel1}) were taken with the neighboring TSU 0.4m T3 APT at Fairborn Observatory (but with different comparison and check stars) and its basic data reduction coefficients are therefore practically identical and the two data sets straightforward to combine.

All data were taken differentially with respect to a nearby comparison star and a check star. A nearby bright star is used for (automatic) navigation. The comparison/check stars were HD152501/HD151179, HD118670/HD118244, and HD113448/HD112805 for V2253\,Oph, IT\,Com and IS\,Vir, respectively. A single measurement is the mean of three readings between the variable and the comparison star. Sky readings through the same 30\arcsec\ aperture as for all other readings are  taken usually between the comparison and the variable star measure to ensure timeliness. The detailed data reduction and Johnson-Cousins transformations for {\sl Amadeus} were described in Granzer et al.  (\cite{granzer01}). We refer the reader to this and subsequent papers as well as references therein.

\subsection{Determining photometric periods}

Throughout this paper, we use the program package \emph{MuFrAn} (Multiple Frequency Analysis; Koll{\'a}th \cite{kollath}) to quantify rotational periods and its errors, which are determined increasing the residual scatter of the least-squares solutions by about 10\% of the precision of the data.  Another package \emph{TiFrAn} (Time Frequency Analysis; Koll{\'a}th \& Ol{\'a}h (\cite{kollath_olah}) allows a time-frequency analysis with several methods if the datasets are long enough in order to check for possible cycles of length of a few years. In this paper we apply a Short-Term Fourier Transform (STFT), numerical details of the procedure are given in Koll{\'a}th \& Ol{\'a}h (\cite{kollath_olah}), including the input of data with variable  (rotational)  modulation and in the presence of seasonal gaps. 
This algorithm was already used for the detection of cyclic behavior of several other active stars (e.g. Ol{\'a}h et al. \cite{olahetal}).

\subsection{Fitting spectral energy distributions (SEDs)}

We determine effective temperatures by fitting ATLAS-9  models (Castelli \& Kurucz \cite{kurucz}) to optical and infrared fluxes. Optical fluxes at Johnson $BV$ and Cousins $I_C$  wavelengths were derived from our light curves for the times of maximum stellar brightness (presumably the least spotted state). In order to take into account the large variability of stars, we assigned an uncertainty of 0.1~mag to each photometric value. 

Near-infrared $JHK_s$ data were taken from the Two Micron All-Sky Survey (2MASS; Skrutskie et al. \cite{2mass}). These data were supplemented by photometry from the Wide-field Infrared Survey Explorer (WISE) satellite. We used the $W1$ band (centered on 3.4\,$\mu$m) photometry from the All-Sky Database (Wright et al. \cite{wright}) but rejected the $W2$-band (4.6\,$\mu$m) measurements because these were saturated for all of our targets.

  \begin{figure}[!tbh]
   \includegraphics[width=8cm]{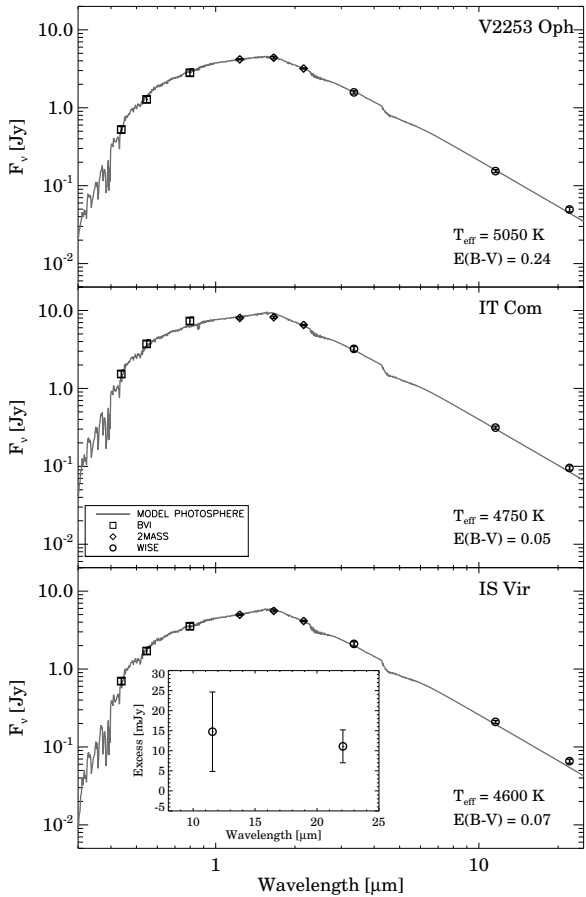}
\caption{Spectral energy distributions of V2253\,Oph (top), IT\,Com (middle) and IS ~Vir (bottom). The dots are the observed fluxes from our APT data, from 2MASS, and from WISE. The lines are the best fit ATLAS-9 models. Effective temperatures and interstellar extinction values are given in each panel. }
   \label{SEDs}
\end{figure}

We fixed both the surface gravity  ($\log{g}$) and the metallicity ([Fe/H]) during the fitting procedure. For IT\,Com and for IS\,Vir these values were set to the literature data cited in Table~\ref{table:1} (Morel et al. \cite{morel}). We then selected   model atmospheres from the ATLAS-9 grid that have the closest metallicity and $\log{g}$ value to the quoted ones. In the case of V2253\,Oph no literature data for gravity and metallicity were found. Therefore, we adopted solar metallicity and estimated the surface gravity from its spectral type and luminosity class  (see Table \ref{table:2}).  All of our targets have good quality photometry in the mid-infrared $W3$ (12\,$\mu$m) and $W4$ (22\,$\mu$m) bands with signal-to-noise ratio $>$20:1. By utilizing the $W3$ and $W4$ bands, we are also able to search for dust emission around our targets. None of our sources are affected by any known artifacts. As a first step, the predicted photospheric flux densities were derived for the WISE wavelengths using the best-fit ATLAS-9 models. Then the significance level of any infrared excess was calculated for both mid-infrared WISE bands by $\chi_{\lambda} = (F_{\rm WISE} - F_{\rm pred}) /\sigma_{\lambda}$, where $F_{\rm WISE}$ is the flux in a specific WISE band and $F_{\rm pred}$ is the corresponding predicted flux value from ATLAS-9.  $\sigma_{\lambda}$ is the quadratic sum of the uncertainty of the measured flux density, the absolute calibration uncertainty (taken from the Explanatory Supplement of the WISE All-Sky Data Release Products), and the uncertainty of the predicted flux density in the specific band. Following the general approach for clear excess detection, we require $\chi_{\lambda} \geq 3.0$ 
(e.g. Kennedy \& Wyatt \cite{kennedy}).

\subsection{Estimating stellar ages}

Ages are estimated from a comparison of the location in the H-R diagram with evolutionary tracks from Pietrinferni et al. (\cite{Pietrinferni}). Stellar luminosity is based on the \emph{Hipparcos} parallax (van Leeuwen \cite{hipparcos}) and the brightest magnitudes of our $V$-band light curves. Bolometric corrections were computed based on $T_\mathrm{eff}$
values using the formula outlined in Torres (\cite{torres}), and we adopted an absolute bolometric magnitude of 4.73 for the Sun. Luminosity errors are simply propagated parallax and magnitude errors. The temperatures and its errors are those from the SED fits.

There is an indirect dynamical method to estimate the age of a binary system, or at least to give some constraints for it. The method is based on a comparison of the present-day orbital configuration with the theoretical tidal circularization time scales. At this point we refer to  Verbunt \& Phinney (\cite{verbuntphinney95}) who investigated the circularization in binaries where the primary component is an evolved, red giant star. From the equilibrium-tide theory of Zahn~(\cite{zahn77,zahn89}) these authors  derived a practical formula (their  Eq.~[7]) which gives the rate of circularization for every evolutionary stage of an evolved primary as a function of the mass-ratio and the orbital period. We note that this formalism depends only weakly on the (uncertain) mass-ratio but strongly on the accepted evolutionary model, and thus on the mass of the primary.

\begin{table}[!tbh]
\caption{Revised stellar parameters.}\label{table:2}
\begin{tabular}{llll}
\hline\hline
\noalign{\smallskip}
Parameter                                           & V2253~Oph & IT~Com & IS~Vir \\
\hline\noalign{\smallskip}
 $V_{\rm max}$                      & 8.61 & 7.45 & 8.30\\
 $V_{\rm min}$                       & 8.87 & 7.86 & 8.70\\
$\langle V-I_C\rangle$           & 1.25-1.35 & 1.12-1.24 & 1.17-1.28\\
$P_{\rm rot}$ (days)                      & 21.55$\pm$0.03  & 65.1$\pm$0.3  &  23.50$\pm$0.04 \\
Cycle?                                            & 10 yr & flip-flop & 5-6 yr, 2 yr\\
$T_\mathrm{eff}$ (K)                     & 5050$\pm200$ &  4750$\pm130$  &  4600$\pm90$   \\
$E(B-V)$                                        & 0.24$\pm0.08$ & 0.05$\pm0.05$ & 0.07$\pm0.04$ \\
$\log L_1$ (L$_\odot$)                  & 1.97$\pm0.37$ & 1.68$\pm0.15$ & 1.66$\pm0.21$ \\
$R_1$ (R$_\odot$)                        & 12.7$\pm$5.4 & 10.3$\pm$1.8 &10.7$\pm$2.6 \\
$M_1$ (M$_\odot$)$^{\it a}$ & $\approx$3 & $\approx$1.4 & $\approx$1.3 \\
$M_2$ (M$_\odot$)                       & 1.35: & 0.29: & 0.55: \\
$i$  (\degr)                                     & $\approx$65 & 40$\pm$12 & 15$\pm$5 \\
$a_1$ ($10^6$\,km)                      &  68.3 &  9.38 & 8.87 \\
\hline\noalign{\smallskip}
\multicolumn{4}{c}{Constant values used when calculating the revised parameters}\\
$\log g$ $^{\it b}$ & 2.5 & 2.5 & 2.5 \\
Fe/H $^{\it c}$ & 0.0 & $-$0.2 & 0.0 \\
distance (pc)$^{\it b}$  & 360$\pm105$ & 185$\pm25$ &  250$\pm50$\\
\noalign{\smallskip}\hline
\end{tabular}
\noindent {$^{\it a}$ Possible mass ranges are given in the text. $^{\it b}$ For IT~Com and IS~Vir  Fe/H is from the literature, for IS~Vir  we suppose solar abundance because of the contradicting (but close to the solar) literature values. The $\log g$ of IT~Com of Soubiran et al (\cite{soubiran}) has no published error and is considered uncertain. $^{\it c}$From the corrected Hipparcos parallax. }
\end{table}

\begin{table}[h]
\caption{Newly determined galactic coordinates and velocities}            
\label{table:3}     
\begin{tabular}{l c c c}   
\hline\hline                 
par. (unit)& V2253~Oph & IT~Com & IS~Vir \\    
\hline                        
v$_{rad}$(km/s)$^{\it a}$ & $-$36.75$\pm$0.13 & $-$19.13$\pm$0.08 & 21.25$\pm$0.04\\
U (km/s)& $-$34.9$\pm$0.7 & $-$63.8$\pm$9.0 & 12.4$\pm$1.5 \\
V (km/s)& $-$13.8$\pm$7.1 & $-$88.6$\pm$13.2 & $-$26.8$\pm$4.2 \\
W (km/s)& $-$19.9$\pm$5.7 & $-$5.3$\pm$2.1 & 7.2$\pm$2.5 \\
X (pc)& 352.1 & 37.8 & 85.1 \\
Y (pc)& $-$29.9 & 1.9 & $-$101.5 \\
Z (pc)& 67.5 & 180.6 & 210.5 \\
b (deg.)$^{\it b}$& 10.81 & 78.18 & 57.82\\
\hline   
\end{tabular}
\vspace*{2mm}

\noindent {$^{\it a}$Gamma velocities form the spectroscopic orbits. $^{\it b}$Galactic latitude. In the calculation of the Galactic space velocity we used a right-handed coordinate system (U is positive toward the Galactic center, V is positive in
the direction of Galactic rotation, and W is positive toward the north Galactic pole) and followed the general recipe described in the Hipparcos and Tycho catalogs (ESA 1997). X, Y, Z are the physical space coordinates, centered on the Sun,
in the same directions as U, V, W.For the calculation of galactic space motion components the coordinate, proper motion and trigonometric distance information were taken from the $Hipparcos$ catalogue (van Leeuwen \cite{hipparcos}).} 
\end{table} 

\section{Masses and ages}\label{S4}

\begin{figure}[!tbh]
{\bf a)}\\
   \includegraphics[width=8cm,clip]{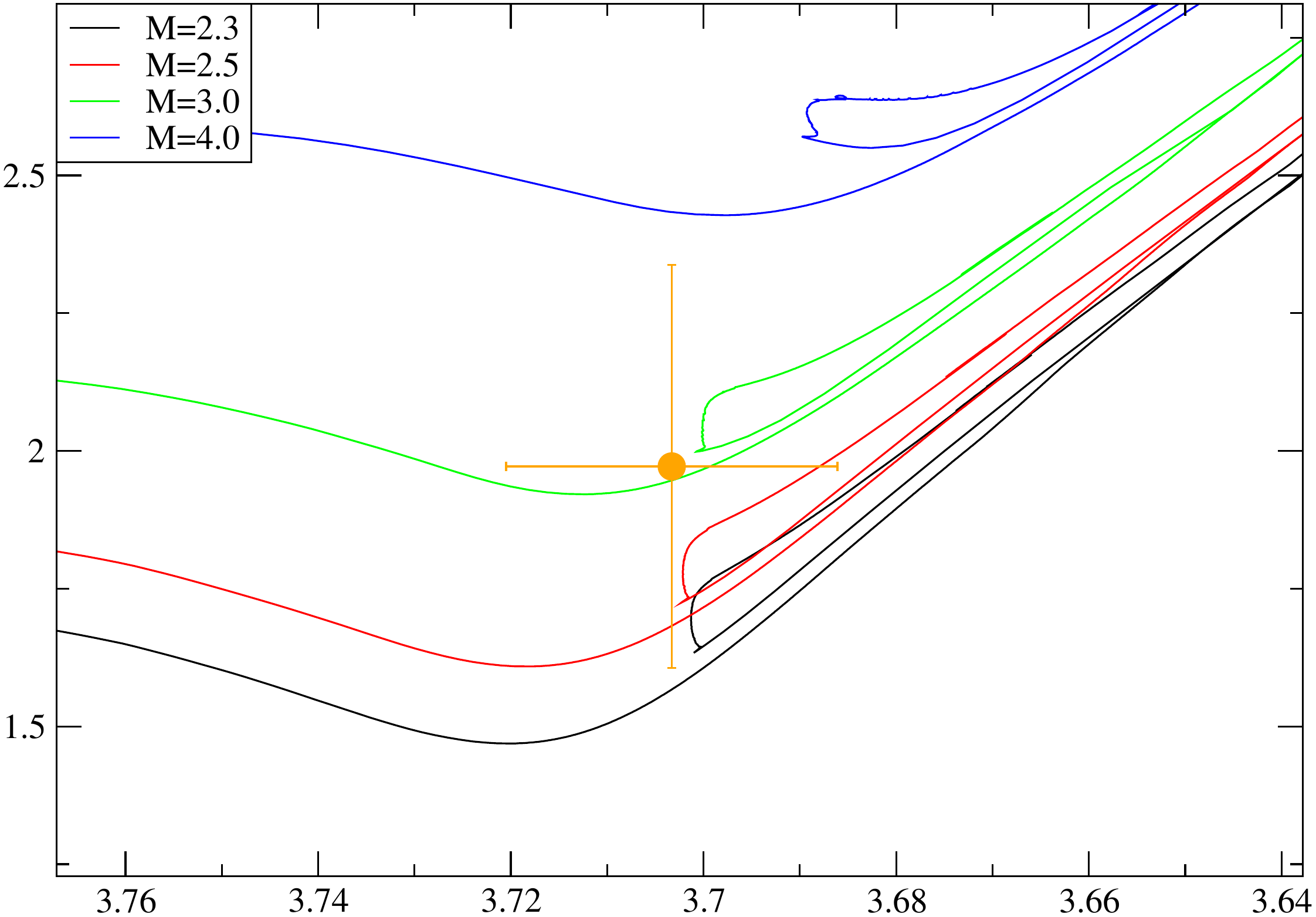}\\
   
{\bf b)}\\
    \includegraphics[width=8cm,clip]{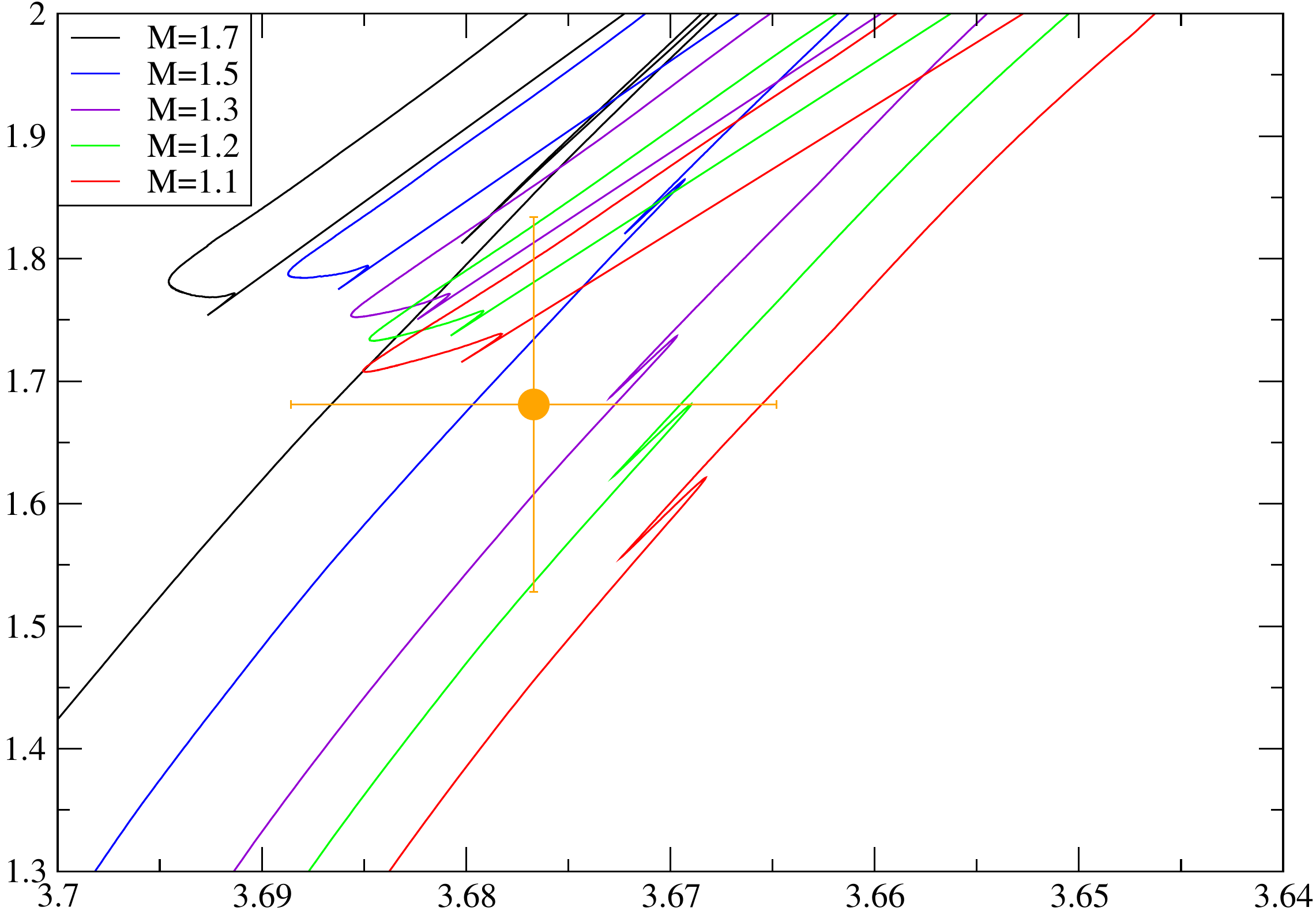}\\
    
{\bf c)}\\
   \includegraphics[width=8cm,clip]{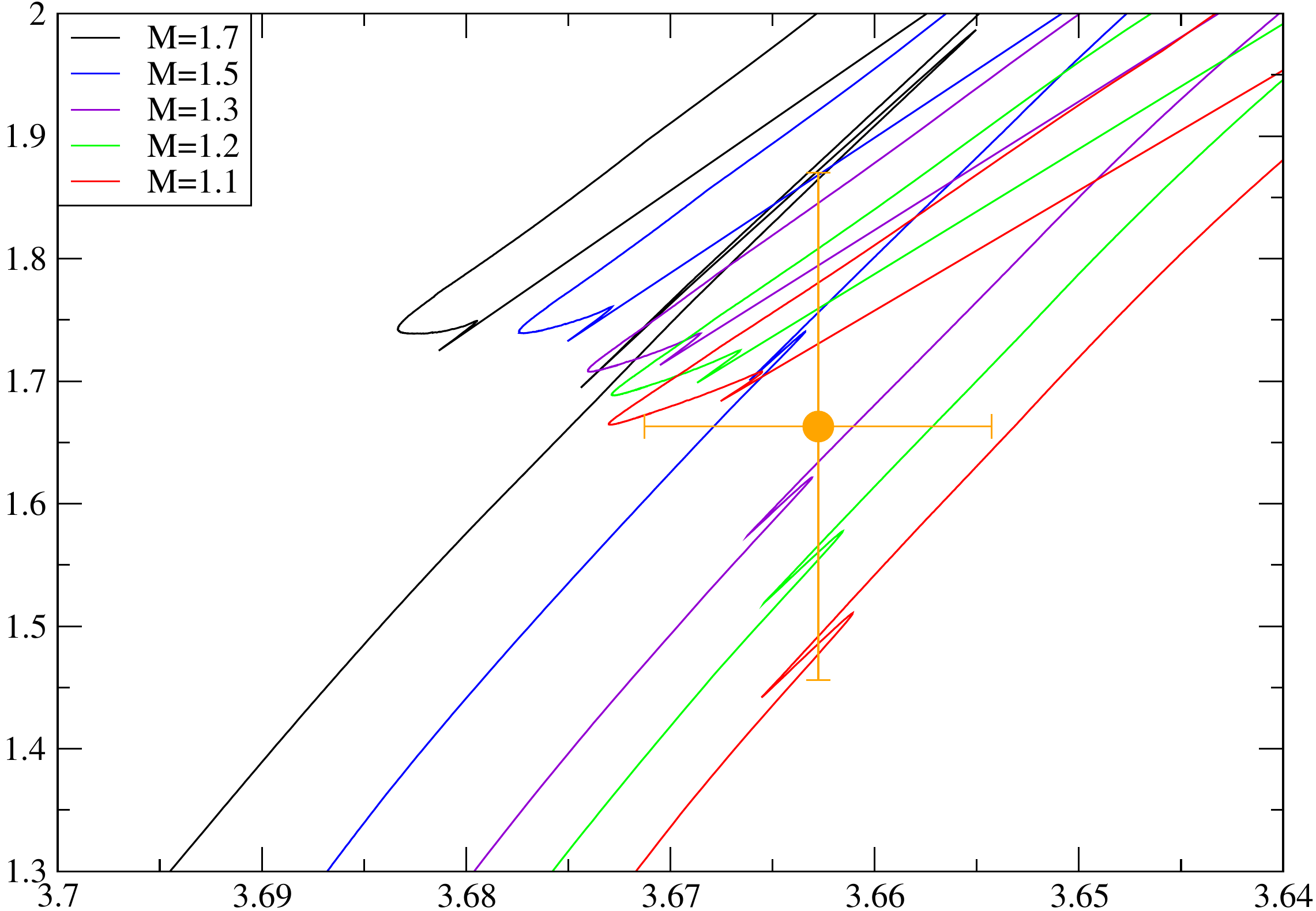}
   \caption{Positions of the three primary stars in the H-R diagram (dots with error bars). From top to bottom,  \emph{a.} V2253~Oph,  \emph{b.} IT~Com, and \emph{c.} IS~Vir. The vertical axis plots logarithmic luminosity in solar units and the horizontal axis is logarithmic effective temperature in Kelvin. The lines are tracks for different masses as indicated in the inserts and were computed for solar metallicity for V2253~Oph and IS~Vir, but interpolated to [Fe/H]=$-0.2$ for IT~Com. }
         \label{HRD}
\end{figure}

The spectral energy distributions (SEDs) of V2253~Oph, IT~Com and IS ~Vir are plotted in Fig.~\ref{SEDs} and the resulting effective temperatures and extinction values are listed in Table~\ref{table:2}. Detailed H-R diagrams are given in  Fig.~\ref{HRD}.

\subsection{V2253~Oph}

For V2253\,Oph, we found the near-IR and mid-IR fluxes to be consistent with pure photospheric emission of $T_\mathrm{eff}$ = 5050$\pm200$K. This value is close to the recent determination of Bailer-Jones (\cite{Bailer}) of 4944$^{+293}_{-336}$ and 4814$^{+174}_{-251}$K, using BVJHK photometry and parallax with two different methods. Our quite large E(B-V)=0.24$\pm0.08$ is also supported by Bailer-Jones' (\cite{Bailer}) values of 0.18$^{+0.11}_{-0.12}$ and 0.14$^{+0.06}_{-0.10}$ from the two methods, respectively, Errors mark the 90\% upper- and lower bounds for both parameters.

The position in the H-R diagram in Fig.~\ref{HRD}a shows that V2253~Oph is quite massive, about 3~M$_\odot$ with a mass range between 2.4--3.8~M$_\odot$. Hence, the system cannot be very old, maybe 1~Gyr or even less according to the models of Pietrinferni et al. (\cite{Pietrinferni}).

From the formula of Verbunt \& Phinney~(\cite{verbuntphinney95}) we find that in case of the $\approx$314-d orbital period,  we could expect a circularized orbit only as late as during the AGB phase of stellar evolution.
Apparently, the star is not in a such late phase of its evolution. A possible solution is offered by Verbunt \& Phinney~(\cite{verbuntphinney95}). They suggest, that for some circularized very wide binaries, a  possibility to explain the circular orbit with such a massive primary is, that the unseen secondary could be a white dwarf. It is thought that the white dwarf forces the circularization during the primary's red giant phase either due to the stronger tidal friction during the epoch when the primary radius becomes very large or/and a more rapid mass-transfer in case of a Roche-lobe filling configuration. 

The orbit of the V2253 Oph system is fairly wide, so the suggestion of Verbunt \& Phinney~(\cite{verbuntphinney95}) could be considered. The minimum radius of the primary of 12.7~R$_\odot$ together with the rotational period of 21.55\,d and a measured  $v\sin i$  of 28.8~\kms\ (Fekel et al. \cite{fekel2}) and 24~\kms\  (Strassmeier et al. \cite{Ca_sp}) yield moderately high inclinations of the rotational axis of $i\sim70\degr$ and $54\degr$, respectively for these $v\sin i$ values. Assuming a coplanar orbit, this suggests a mass ratio between q$\sim$0.43--0.51 resulting in a secondary with a mass of about 1.3~M$_\odot$ orbiting the 3~M$_\odot$ primary. On the other hand, if the primary mass is just 2.5~M$_\odot$, then the secondary mass would be slightly less than 1.2~M$_\odot$. More about the mass of the secondary is in Sect.~\ref{A2} in the appendix.

The existence of such a high mass white dwarf has been proven by Kalirai et al. (\cite{kalirai}), who studied white dwarfs
in the young (650~Myr) cluster NGC~2099. For WD~24 a present-day mass of 1.11~M$_\odot$ was claimed and an initial mass of 4.43~M$_\odot$ computed, although with considerable errors. Thus, a scenario of V2253~Oph with a red-giant primary of 2.4~M$_\odot$ and a white-dwarf secondary of about 1.1~M$_\odot$, which should have been originally over 4~M$_\odot$, is not impossible,  taking into account the young age of the system, which could have been still long enough for the present secondary to evolve to a high-mass white dwarf.

\subsection{IT~Com}

For IT~Com, we found the measured mid-infrared fluxes to be consistent with pure photospheric emission of $T_\mathrm{eff}$ = 4750$\pm130$K. Bailer-Jones' (\cite{Bailer}) results of 4821$^{+337}_{-333}$ and 4735$^{+237}_{-213}$K with two different methods are practically the same as our result, although with quite high uncertainty. The near zero E(B-V)=0.05$\pm0.05$ we found is lower than that of Bailer-Jones's (\cite{Bailer}) of 0.14$^{+0.12}_{-0.13}$ and 0.11$\pm$0.09 from the two methods, but still are well within the respective errors.

The position in the H-R diagram suggests a comparable low mass of 1.4~M$_\odot$  (Fig.~\ref{HRD}b) with a full range of 1.2--1.6~M$_\odot$. The Pietrinferni et al. (\cite{Pietrinferni}) tracks indicate an age of $\approx$1.6 Gyrs.  Large dispersion in the galactic space velocities of IT~Com (Table~\ref{table:3}) implies that it may not be a thin disc star. Indeed, using the kinematic criteria proposed by Reddy et al. (\cite{reddy}) we found that IT~Com probably belongs to the thick disc population.

The orbit has a high eccentricity  which, together with the dynamic age based on Verbunt \& Phinney (\cite{verbuntphinney95}), suggests that IT~Com is in some stage of H-shell burning and not yet on the AGB because then the orbit should have already been circularized.

The calculated radius of 10.3~R$_\odot$ and $v\sin i$ of 5.5$\pm$1~\kms\  (Henry et al. \cite{henry_etal}) constrain the inclination of the rotational axis to between 34--54$\degr$, in good agreement with the most likely coplanar orbital inclination of about 40$\degr$, derived in Sect.~\ref{A1} in the appendix. The mass of the secondary star is about 0.55M$_\odot$, see Sect.~\ref{A2} in the appendix.

\subsection{IS~Vir}

For IS~Vir, we derived $T_\mathrm{eff}$ = 4600$\pm90$K, very similar to the recent determination of Bailer-Jones (\cite{Bailer}) of 4521$^{+155}_{-410}$ and 4574$^{+106}_{-161}$K. Our  E(B-V)=0.07$\pm0.04$ is practically the same as that of Bailer-Jones' (\cite{Bailer}) of 0.03$^{+0.07}_{-0.03}$ and 0.06$^{+0.04}_{-0.06}$ from the two methods.

The SED fits for the mid-IR bands achieved $\chi_{W3} = 1.5$ and $\chi_{W4} = 2.7$, implying a marginal excess detection in the 22-$\mu$m $W4$ band. Recently, a comparable amount of excess emission at $W4$  was announced for HD\,139357, a planet hosting K4\,{\sc iii} giant star (Morales et al. \cite{morales}). Generally, Jura (\cite{jura}) explained excess  IRAS 25\,$\mu$m emission by dust release during ice sublimation of Kuiper belt-like objects as the star evolves from the main sequence onto the red giant branch. Further infrared observations are needed to confirm the excess and to clarify whether it is related to circumstellar dust or can be explained differently, e.g. by the contribution of a background galaxy or by an apparent positional coincidence with an interstellar dust cloud. 

The position of IS\,Vir in the H-R diagram is compatible with a mass of 1.3~M$_\odot$ (Fig.~\ref{HRD}c), i.e. of similar mass than IT~Com. The mass range is 1.1-1.5 and the age of the system is certainly beyond 5 Gyrs from the HRD.

Fekel et al. (\cite{fekel1}) derived a similar mass of 1.5~M$_\odot$ and claims the system is not young since the star is about 250~pc above the galactic plane. The tidal circularization time scale does not provide any further constraints for its age because the orbital-period limit for circularization is longer or comparable (28\,d) to today's orbital period (23.6\,d).

IS~Vir has a low orbital inclination of about 13$\degr$ according to (Fekel et al. \cite{fekel1}). From its newly calculated radius of 10.7~R$_\odot$, its rotational period of 23.5\,d, and the projected rotational velocities found in the literature.  The inclination lies between 10--20$\degr$; the range of inclination comes from the uncertainties of the radius and the different velocity data collected in the CABS-III catalog (Eker et al.~\cite{CABSIII}). The star is seen nearly pole-on. As a consequence of the low inclination the amplitude of the rotational modulation is always small and less than 0\fm1 in $V$, as can be seen in Fig.~\ref{lightcurves}c. The mass of the secondary component must be $\approx$0.55~M$_\odot$, see Sect.~\ref{A2} in the appendix, for more.

\section{Starspot activity}\label{S5}

\subsection{V2253~Oph}

\begin{figure}
   \includegraphics[width=8cm]{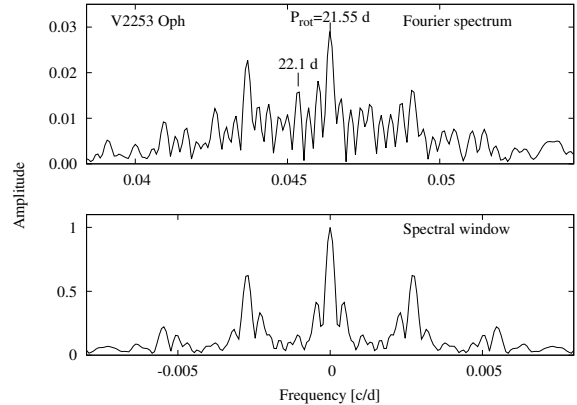}
   \caption{\emph{MuFrAn} periodogram of V2253~Oph from the $V$-band data until the end of 2005. Two significant periods are marked in the upper panel, the spectral window is plotted below.}
         \label{V2253Oph_rotper}
   \end{figure}

The $V$-band amplitude due to rotational modulation varied between 0\fm15 in 1997 at a time of faint average light and 0\fm10 in around 2002 at a time of bright average light (Fig.~\ref{lightcurves}a). This suggests a relation between the overall spottedness and the size of individual spots or spot groups similar to what has been seen on other over-active stars. However, the shape of the light curve of V2253\,Oph changes continuously and smoothly between subsequent rotations. This is indicative of rapid spot rearrangements typically observed for the more active stars.  Not surprisingly, the \emph{MuFrAn} periodogram reveals two closely-space periods, but only until the end of 2005, see Fig.~\ref{V2253Oph_rotper}. The  dominant 21.55$\pm$0.03-d period with an average 0\fm06 full $V$  amplitude is seconded by a 22.05$\pm$0.04-d period with a full 0\fm025 amplitude. We tentatively interpret this due to differential surface rotation and estimate a relative shear factor $\delta P/P$ of $\approx$0.02, i.e. a lap time of the equator with respect to the poles of $\approx$314\,d.   

\begin{figure}
   \includegraphics[width=7cm]{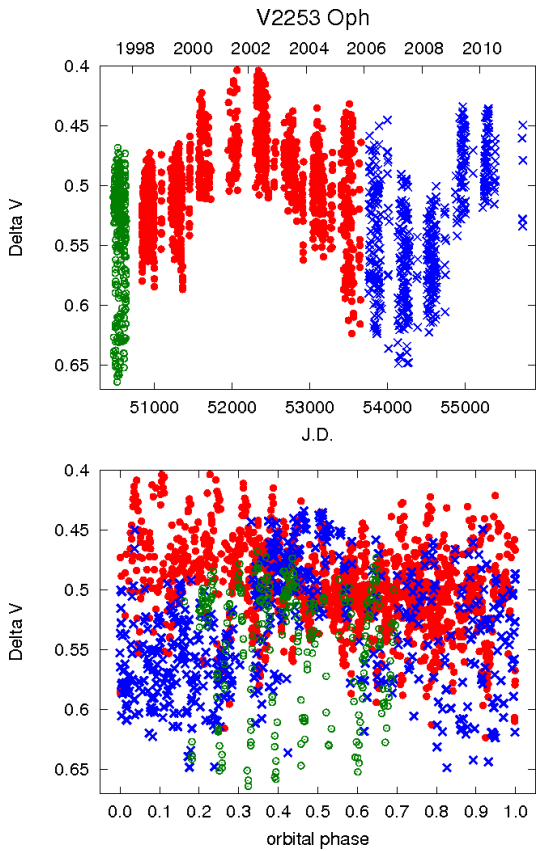}
   \caption{Evidence for a relation of the spot activity on the primary of V2253~Oph and the binary orbit. The top panel shows again the long-term $\Delta V$ magnitudes plotted versus Julian date while the bottom panel shows the same data plotted versus \emph{orbital} phase ($P_{\rm orb}$ = 314.17\,d). Symbols on both panels refer to the same data points.}
         \label{V2253Oph_orbph}
\end{figure}

   \begin{figure*}
{\bf a} \hspace{80mm}{\bf b}  \\
   \includegraphics[width=8cm]{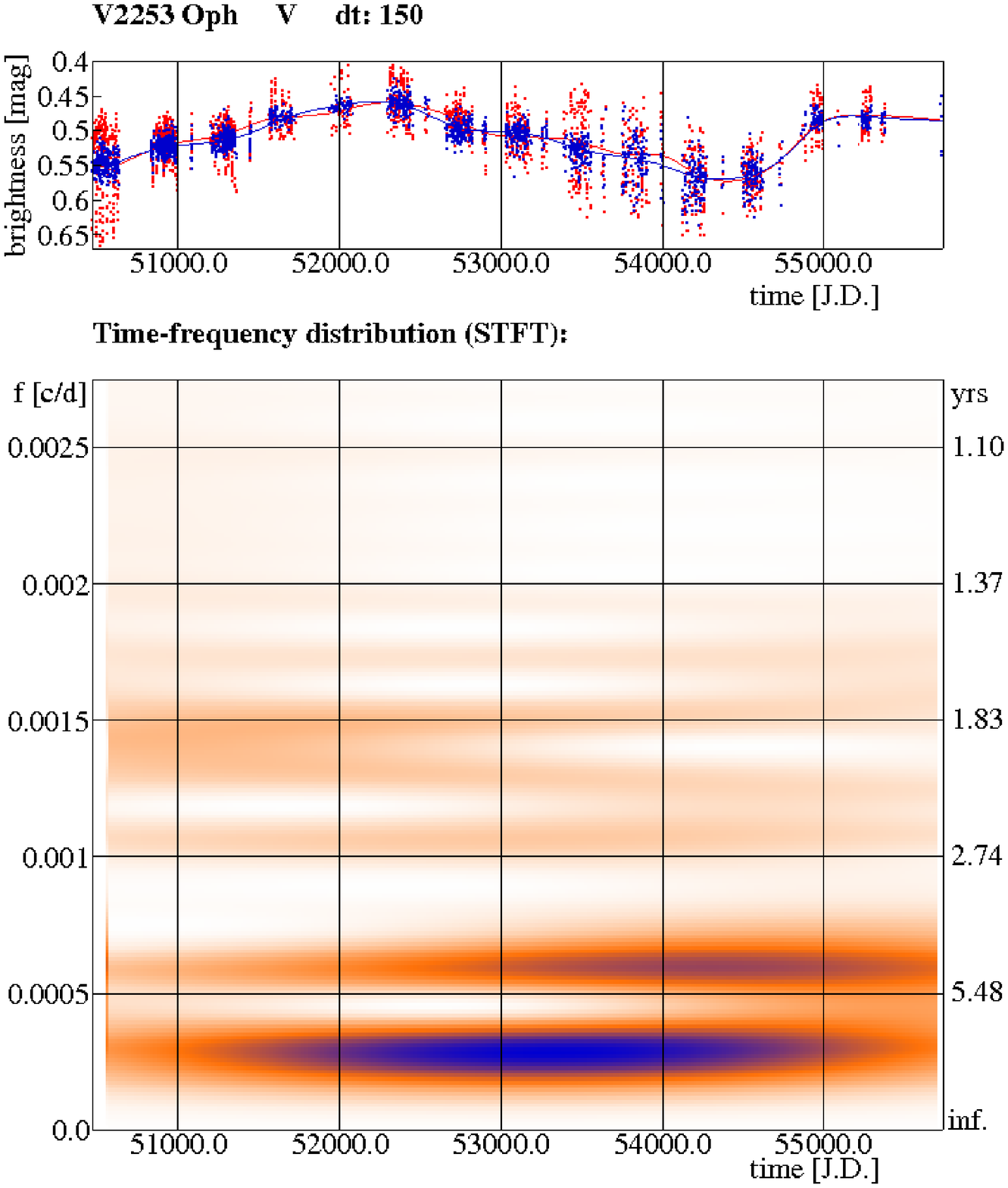}\hspace{5mm}
   \includegraphics[width=8cm]{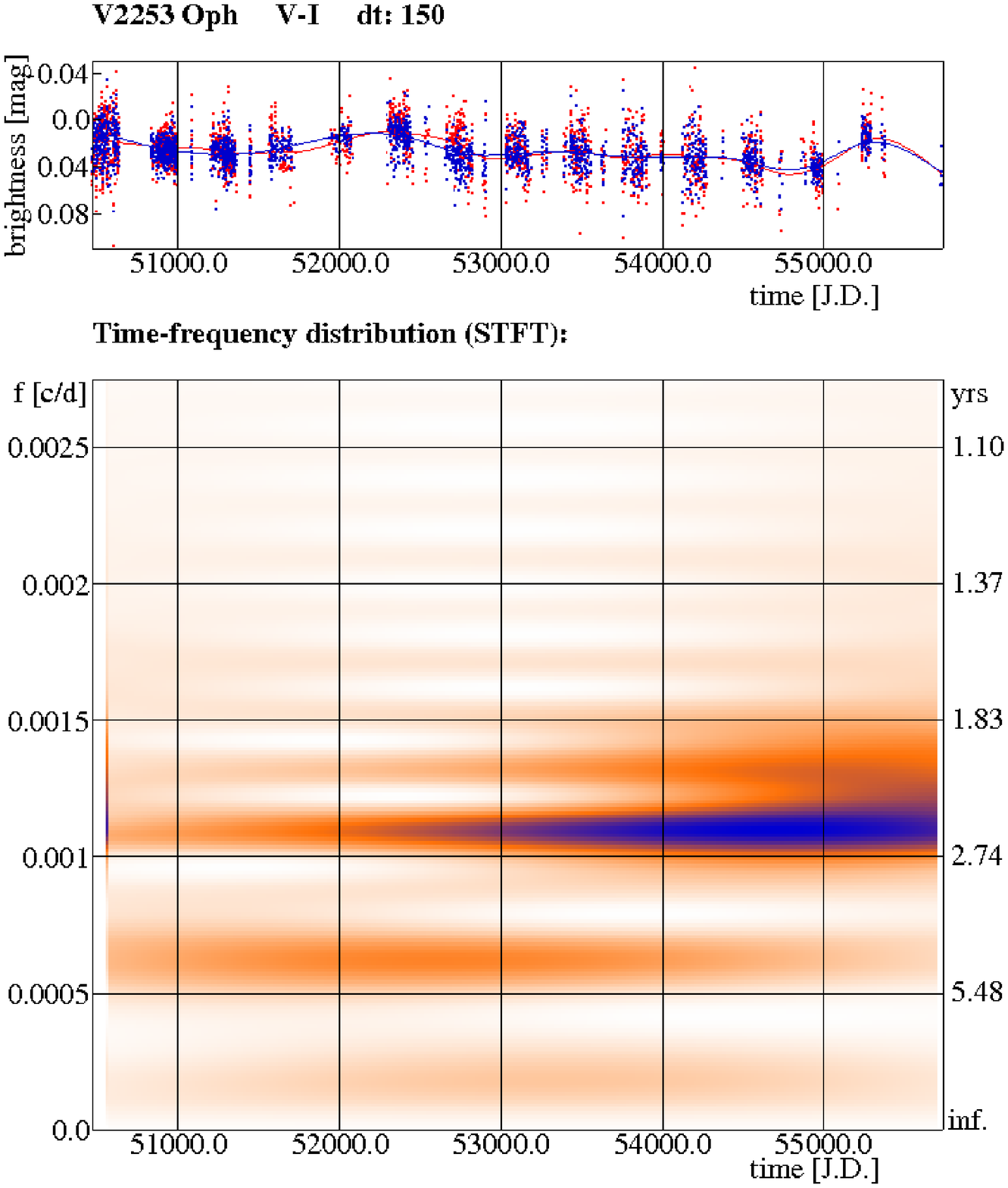}
   \caption{Time-frequency diagrams for V2253~Oph. \emph{a.}  from the $V$ magnitudes and, \emph{b.},   from the $V-I_C$ color index. The top panels show the observations as red (grey) dots, blue (black) dots show the data prewhitened with the main rotational period and its harmonics, red and blue (grey and black) lines show spline interpolations to the respective data. Notice the strongest signals are at around 10\,yrs from $V$ and $\approx$2.5\,yrs from $V-I_C$. }
         \label{V2253Oph_longterm}
   \end{figure*}

The rotation of the spotted primary is strongly asynchronous with respect to its  314-d orbital period. Note that the  eccentricity is very low and does not exceed its 3\,$\sigma$ uncertainty, and can not account for the period difference. Yet the light variation of two subsets of our APT data seem to produce a fairly homogeneous light curve when plotted versus the orbital period (Fig.~\ref{V2253Oph_orbph}).  Its amplitudes are marginally larger than the ``scatter''  but show extrema (maximum and minimum) always near periastron. The very first seasonal data set from 1997, which shows the highest rotational-modulation amplitude of all of our data, appears to have a minimum at orbital phase near 0.5, half between the formal times of periastron and apastron (note that one observing season usually covers 140 nights, i.e. one half of the orbital period). Possibly, the 1997 season belonged to yet another orbit-induced variability pattern. Due to the small eccentricity it is not very likely that this variation originates from magnetic interaction of the two binary components at periastron but nevertheless there remains some evidence for a relation of the spot activity on the primary with the orbit of the system.

The time-frequency diagram from 15 years of photometry of V2253\,Oph is plotted in Fig.~\ref{V2253Oph_longterm}. The most prominent long-term variation is even seen by eye and stretches over 10~yrs with a peak-to-peak $V$ amplitude of 0\fm1. Despite that the time-frequency diagram gives a strong signal around 10 years, we can not be sure of its cyclic nature due to the relatively short time coverage. Half of this period, a variability of  about 4.5 years, also appears in the time-frequency diagram. The similarly constructed plot from the $V-I_C$ color index does not show any traces of the dominant 10-yr variation but shows a weak signal around 4.5 years. However, the largest amplitude in the  $V-I_C$ color index variation is reconstructed on the timescale of 2.5 years.  All these periodicities seem to be related to each other because within their uncertainties of about 0.5--1\,yr they appear to be harmonics. Note that the color scales in the two plots in  Fig.~\ref{V2253Oph_longterm}  are the same, which allows to directly compare the amplitude of the signals.  We also note that despite the scatter of the measurements was larger between 2007--2009 due to an instrumental problem, the comparison-check data proofed that the mean light level was not affected, so we conclude that the large long-term variation is real.

Unfortunately, we are left with an inconclusive situation. It is tempting to interpret the long-term $V$-brightness variation due to a dynamo-induced change of the overall spottedness, like for the solar 11-year cycle, and the long-term $V-I_C$ index variation due to an average spot-temperature change. However, the two tracers should vary with the same ``cycle'' period or, alternatively, are not related with each other. A bit frustratingly we conclude that after 15 years of photometry we need another 15 years of photometry.

\subsection{IT~Com}

   \begin{figure}
   \includegraphics[width=8cm]{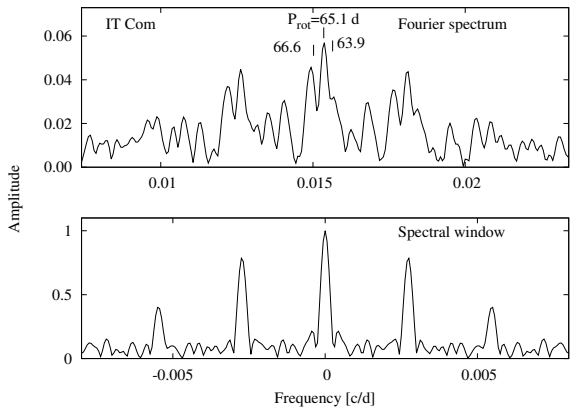}
   \caption{\emph{MuFrAn} periodogram of IT~Com from the $V$-band data. Three significant periods are marked in the upper panel, the spectral window is plotted below.}
         \label{ITCom_rotper}
   \end{figure}

\begin{figure}
   \includegraphics[width=7cm]{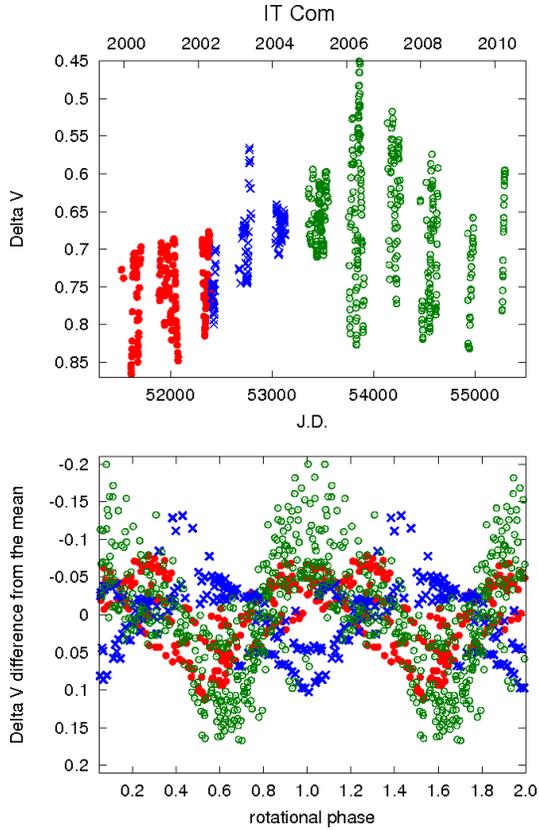}\\
   \caption{\emph{Upper panel:} $\Delta V$ magnitudes of IT~Com versus JD for 2000--2010. The plot symbols and colors mark the data embraced by the flip-flop. The first phase flip started in early 2002 and flopped back by early 2005. \emph{Lower panel:} The same data as above but phased with the rotational period of 65.1\,d. The plot symbols and colors are the same as in the top panel and indicate the phase shift of the data within the flip flop. Note that the lower panel normalized the differential magnitudes to the mean light level of the three subsets. }
         \label{ITCom_flipflop}
\end{figure}

The best-fit average photometric period for IT~Com from the 11-yr $V$-band data is 65.1$\pm$0.3\,d (Fig.~\ref{ITCom_rotper}). However, the photometry shows significant light-curve changes from one stellar rotation to the next (note that only two stellar rotations can be covered during one observing season) that may affect its period determination. An amplitude increase of the rotational modulation by nearly a factor of three happened between 2005 and 2006.  In 2002, a sudden phase shift by $\approx$0\fp4 was witnessed during the observing season which then reverted back to the original phase in early 2005. This could be termed a classical flip-flop phenomenon (e.g. Elstner \& Korhonen \cite{els:kor}, Ol\'ah et al. \cite{olah-ff}). It  has started near the periastron passage of the two binary components at JD 2,452,406.734, marked in Fig.~\ref{ITCom_flipflop} top panel  where the observations before, during, and after the phase flips are plotted with different symbols (and color).  The back-flip in 2005 was actually not directly seen because it happened outside the annual observing window but the phase coherence thereafter was the same as before the flip and constant until the end of the APT observations in 2010.  The time duration from flip to flop was between 2.0--2.7 years, i.e., it ended between the last observation of 2004 and the first one in 2005. So far, the phenomenon did not repeat and we could not speak of a flip-flop ``cycle'' but state that flip flops may occur aperiodically and its length is not related to its reoccurrence.

\begin{figure}
   \includegraphics[width=8cm]{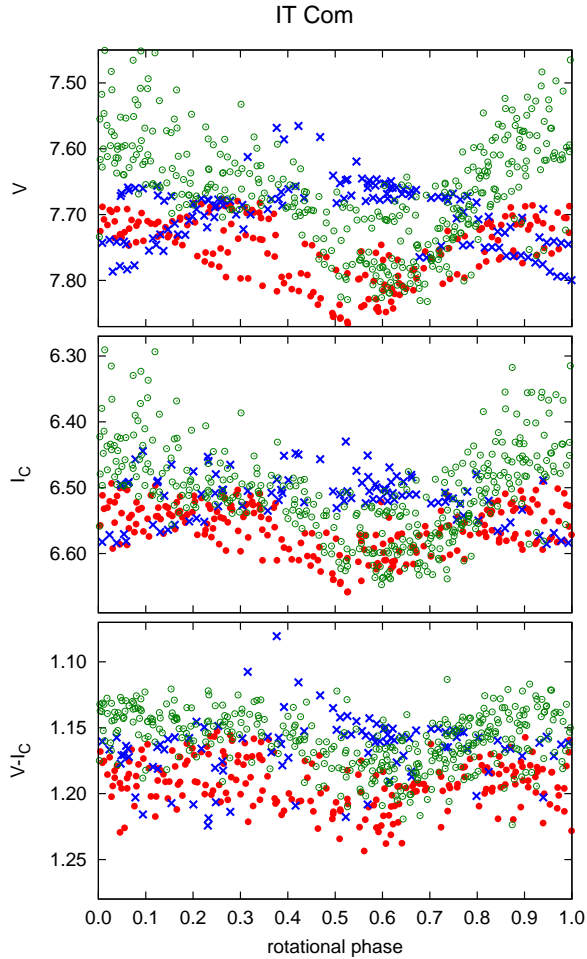}
   \caption{Phased light and color index curves of IT~Com with the 65.1 days period, phase zero is at periastron. The plot symbols are the same as in Fig.~\ref{ITCom_flipflop}. $V-I_C$ is redder when the star is fainter. }
         \label{ITCom_ph}
   \end{figure}

Systematic seasonal changes are also obvious for the average $V$-light level in Fig.~\ref{lightcurves}b (replotted on a larger scale in  Fig.~\ref{ITCom_flipflop}, top panel) and amount to almost 0\fm15. Unfortunately, the 11 years of photometry is insufficient to search for cyclic changes; a simple look at the data suggests a timescale of a cycle (if any) longer than the length of the dataset itself. The long-term $V-I_C$ light curve traces the average $V$-light level in the sense that  redder $V-I_C$ occurs when the system is brighter. The $V-I_C$ variation is in phase with the color variability, at the same time the amplitude of the rotational color modulation does not relate to the amplitude of the $V$ (or $I_C$) modulation, which indicates a constant spot temperature, see Fig.~\ref{ITCom_ph}. 

Apart from the dominant rotational signal of 65.1\,d (with 0\fm11 full $V$ amplitude), two more significant periods show up, see Fig.~\ref{ITCom_rotper}.  One is slightly longer, 66.6\,d with an amplitude of 0\fm072, and one slightly shorter, 63.9\,d with an amplitude of 0\fm068.  Both periods have an approximate error of $\pm$0.3\,d and are highly significant. 

Between 2000--2002 the 65.1-d period is dominant (red dots in Fig.~\ref{ITCom_flipflop}), whereas during the next two years the 66.6-d period gave the best fit (blue crosses in Fig.~\ref{ITCom_flipflop}). From 2005 onwards again the 65.1-d period was dominant and resulted in the least scattered light curve. This indicates that the individual periods stem not from longitudinal migration due to some sort of a differentially rotating surface but are related with the emergence, or decay, of an active longitude as part of the flip-flop phenomenon. If we nevertheless assume a solar-like differential rotation law, adopt a latitude range for the spots of 50$\degr$, and assume the shortest period belongs to the equator and the longest to a latitude of 50$\degr$, we may estimate a lower limit for the relative differential rotation of $\delta P/P\approx 0.07$.  Therefore, the lap time of the equator with respect to the high latitudes is $\approx$90\,d.

\subsection{IS~Vir}

   \begin{figure}[h]
   \includegraphics[width=8cm]{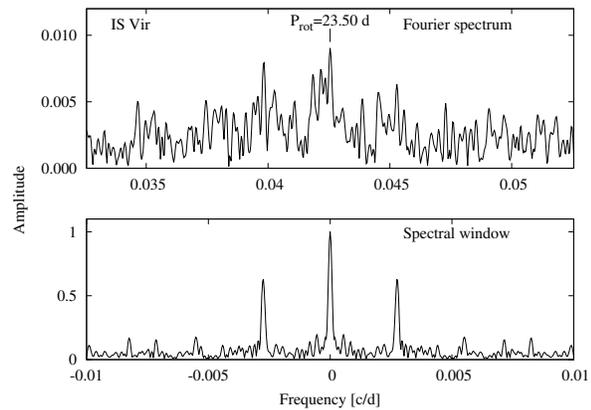}
   \caption{\emph{MuFrAn} periodogram of IS~Vir from the $V$-band data. One period is marked in the upper panel, the numerous peaks around it indicate other, slightly different periods. Note the low amplitude of the variations, see the text for more. The spectral window is plotted below.}
         \label{ISVir_rotper}
   \end{figure}
   
   \begin{figure}
   \includegraphics[width=6.7cm, angle=270]{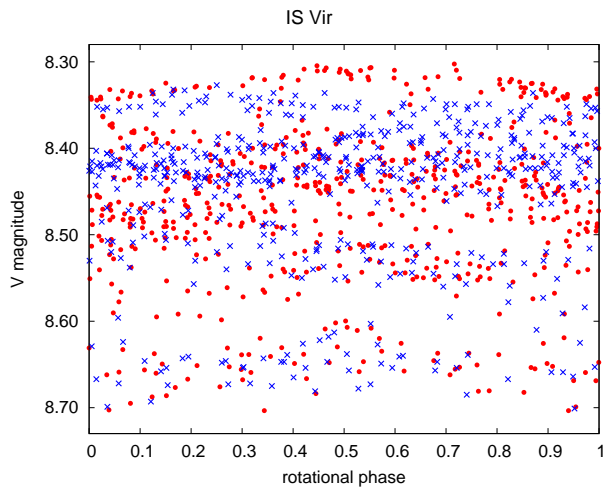}
   \caption{Phased light curves of IS~Vir with the 23.50 days rotational period. Red and blue (dots and crosses) symbols mark our observations and that of Fekel et al. (\cite{fekel1}). Note the very low or sometimes zero amplitude of the rotational modulation and the long-term brighntess change of about 0.4 magnitudes.}
         \label{ISVir_phase}
   \end{figure}

   \begin{figure}[h!!!]
   \includegraphics[width=8cm]{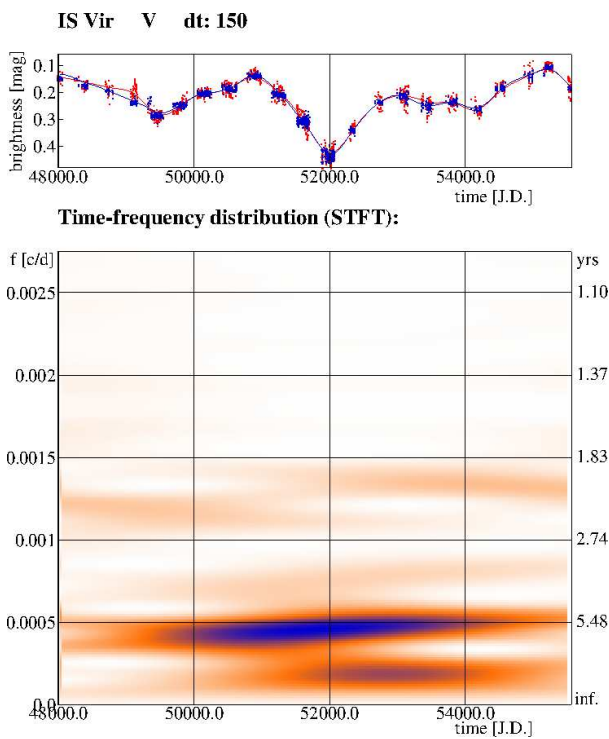}
   \caption{Time-frequency diagram for IS~Vir. The top panel shows the combined data from the TSU T3 APT and the Vienna-Potsdam Amadeus APT.  Observations are plotted as red (grey) dots, blue (black) dots show the data prewhitened with the main rotational period and its harmonics, red and blue (grey and black) lines show spline interpolations to the respective data.}
         \label{ISVir_longterm}
   \end{figure}

Fekel et al. (\cite{fekel1}) published 12 seasons of $BV$ photometry from the TSU T3 APT of which the last two seasons overlap with our first two of the 11 seasons of Vienna-Potsdam APT data. Based on the comparison of the 76 common nights during the two overlapping seasons, we combined the two datasets with a shift of 2.07 mag. in $V$ (due to two different comparison stars) to a 21-year time series for further analysis.  After pre-whitening with the long-term high-amplitude variations with a 6250 days period and its four harmonics (which is a purely mathematical tool to get rid of the long-term variation) we find a mean photometric period of 23.50$\pm$0.04\,d from Johnson $V$ that we interpret to be the stellar rotation period. It is very close but not equal to the 23.655$\pm$0.001\,d orbital period (Fig~\ref{ISVir_rotper}). 

During the first three and the last three seasons when the star was the brightest, the analysis does not give clear periods. The data at medium brightness, i.e., without the first and last three seasons and the season of the minimum, results in two closely-spaced periods of 23.64 and 23.50 days with an error of 0.3 days and a full Fourier amplitude of 0.024 mag. Finally, the year of minimum light shows a photometric period of 22.87$\pm0.80$ days with a full Fourier amplitude of 0.028 mag. Considering the low inclination of the star ($\approx15\degr$) this result suggests, that at maximum light only the polar region has some spots which cause no or very small rotational modulation, while during other times spots may rotate in and out of view at lower latitudes, with shorter period and higher amplitude at minimum. However, according to the errors of the periods and the very low amplitudes we consider this only as a possible scenario. The amplitude change is also seen in $V-I_C$  (Fig.~\ref{lightcurves}c) which suggests that at least part of the long-term variations is due to a temperature change rather than an increase or decrease of the overall spottedness. Fig.~\ref{ISVir_phase} depicts the very small amplitude rotational modulation and the large change in the overall light level of the star during 21 years of observations (blue crosses: Fekel et al. (\cite{fekel1}), red dots: this paper).

The full APT time series in Fig.~\ref{ISVir_longterm} indicates  a long-term variation with a duration of 5-6\,yrs which is slowly decreasing with time, and varying amplitudes of up to 0\fm3 in $V$. A weakly defined timescale of about 2 yrs also shows a decrease from 2.3 to 2.0 yrs. These cycle lengths are not harmonics of each other, therefore their parallel change suggests a common origin. The same feature, i.e., parallel change of two cycles have already been observed in V410~Tau, EI~Eri and UX~Ari by  Ol{\'a}h et al. (\cite{olahetal}).

\section{Summary and discussion}\label{S6}

We have investigated three active (non-eclipsing) SB1 binaries with early K giants as primaries.  The estimated ages of the three systems are naturally rather uncertain, V2253~Oph is about 1~Gyr old or less, while IT~Com and IS~Vir  seem to be older, over 2~Gyrs, likely a few Gyrs. Without more precise distances these estimates can not be significantly improved at the moment. More realistic ages were found for two other active binary systems with giant components; BE~Psc ($\approx$2.7\,Gyrs, Strassmeier et al. \cite{bepsc}) and CZ~CVn (6--7\,Gyrs, Strassmeier et al. \cite{czcvn}). BE~Psc is an eclipsing binary and therefore the stellar parameters are known much better than in case of SB1 systems, and CZ~CVn is less than 100~pc away with just a small extinction, thus its luminosity is better constrained than that of the systems in the present paper. 

V2253~Oph has the highest mass ($\approx$3\,M$_\odot$) of the three active binaries and is also the youngest. Its orbit is nearly circular but the rotation is very fast and strongly asynchronous, the rotational period is almost 15 times shorter than the orbital one. According to Verbunt \& Phinney (\cite{verbuntphinney95}) this unusual scenario suggests either a white dwarf companion or an initially low eccentricity of the orbit. Apart from the rotational modulation the light variation shows a systematic change for several years during the 15 times longer orbital period with extrema (maximum or minimum) around periastron. The two photometric periods found in the data until 2005 are probably due to differential surface rotation which indicates a relative shear factor of $\delta P/P$ of $\approx$0.02, and a lap time of the equator with respect to the poles of $\approx$314\,d. That this value is nearly identical to the orbital period may be coincidental, we have no explanation for it. The activity has a cycle with a characteristic length of about 5 years and a $V-I_C$ color index, thus temperature, variation of about 2.5 years (i.e., half cycle), plus a long-term variation comparable to the length of the dataset. In the literature only two, barely studied similar systems with giant primaries, low (zero) eccentricity and much shorter rotational than orbital periods, are found: DW~Leo and VX~Pyx, however, in both cases the rotational modulation is found be marginal thus their periods are uncertain (see CABS-III: Eker et al. \cite{CABSIII} for the details).

IT~Com and IS~Vir have similar masses (1.4 and 1.3~M$_\odot$), and their orbits are of similar size (a$_\mathrm{1}$ is  9.38$\times10^6$~km and 8.87$\times10^6$~km, respectively),  however the eccentricities are different: IT~Com has a very eccentric orbit while IS~Vir has nearly circular. Both systems are older than V2253~Oph. IT~Com rotates 9\% slower than its orbit and has a very low mass secondary (about 0.3M$_\odot$) so the tidal interaction is not very effective. On the contrary, IS~Vir which has a higher mass secondary (about 0.55M$_\odot$) thus more effective tidal interaction, is synchronized, and the shorter orbital period is below the period limit of circularization. The primaries have strong magnetic activity and the magnetic braking may slow down the rotation, below even the orbital period in case of IT~Com, where the low mass secondary cannot effectively interact. Strassmeier et al. (\cite{binaries}) studied V383~Ser (HD142680), an SB2 system, the only one which shows some similarity with IT~Com: it has a considerable eccentricity (e=0.3158) and rotates super-synchronously (P$_\mathrm{orb}$=24.534 days and P$_\mathrm{rot}$= 33.4 days), but no long-term photometry is available for further studies.

We directly observed an abrupt exchange of the deeper and shallower minima (stronger and weaker active longitude) of IT~Com in 2002. The new configuration lasted for about two years, after that the previous scenario returned. During the 11 years of observations three photometric periods were always present strongly suggesting a differential rotation with a minimum $\alpha\sim0.07$.
From a combined, 22 years long dataset of IS~Vir we found evidence for a slowly decreasing cycle period between about 6-5 years.

\acknowledgements
The authors are grateful to the referee, N. Vogt, for useful suggestions. KO acknowledges support from the Hungarian Research Grant OTKA-K81421. KGS thanks the German State of Brandenburg for continuous support of APT operation and Lou Boyd for making it technically possible.
This publication makes use of data products from the Two Micron All Sky Survey,
which is a joint project of the University of Massachusetts and the Infrared
Processing and Analysis Center/California Institute of Technology, funded by the
National Aeronautics and Space Administration and the National Science
Foundation. This publication makes use of data products from the Wide-field Infrared Survey
Explorer, which is a joint project of the University of California, Los Angeles,
and the Jet Propulsion Laboratory/California Institute of Technology, also funded by
the National Aeronautics and Space Administration.

\appendix

\section{On the inclinations of IT~Com}\label{A1}

   \begin{figure}[hbt]
   \includegraphics[width=7cm]{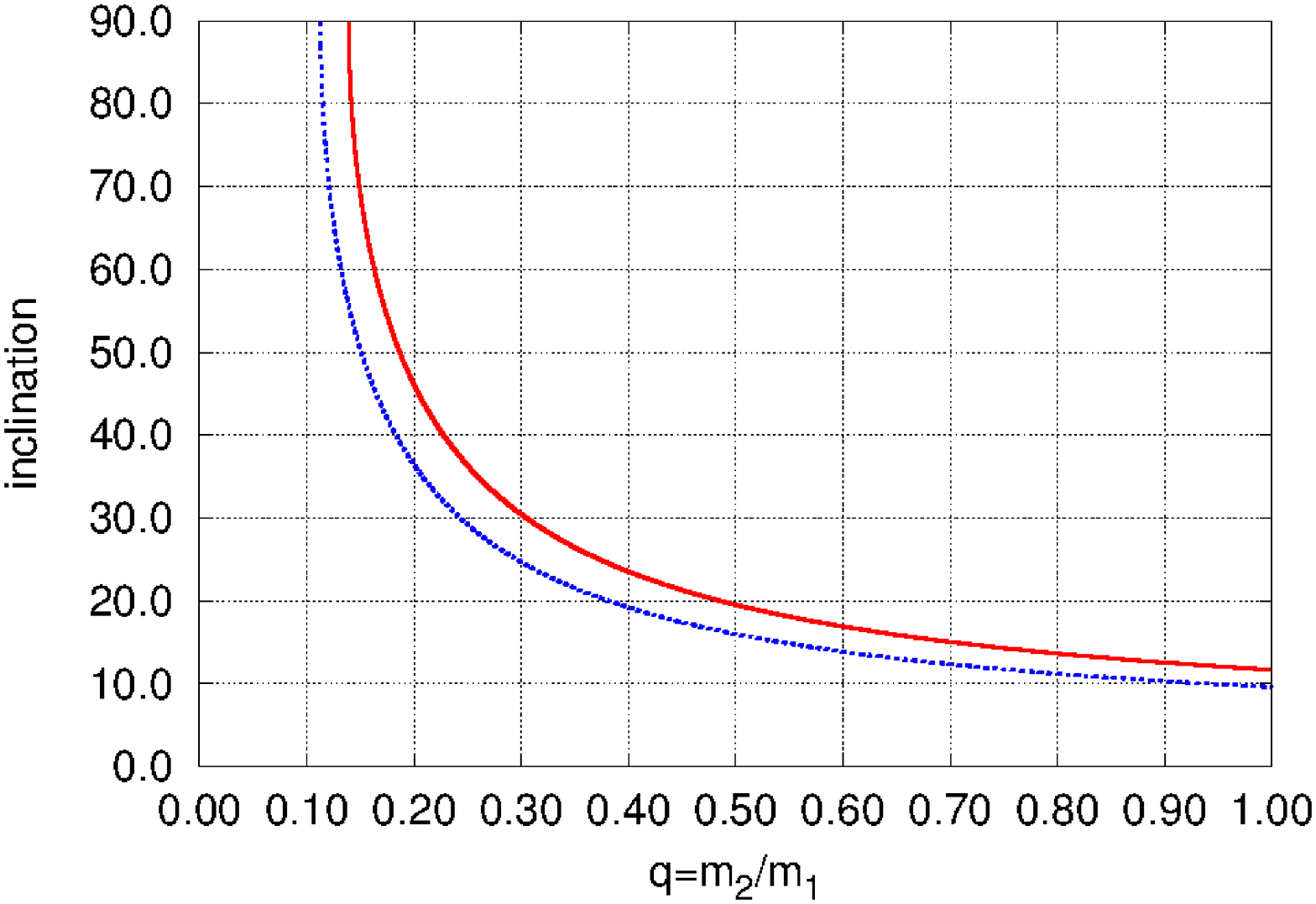}
   \caption{Possible orbital inclinations of IT~Com with different mass ratios. Red solid line and blue dotted line mark primary masses of 1.4 and 2.5 solar masses.}
         \label{incl}
   \end{figure}

The orbital elements of IT~Com were derived by Griffin (\cite{griffin88}), who found a very eccentric orbit ($e$=0.589) and a small mass function ($f(m)$=0.0029). Griffin suggested that the primary was a K0 or a K1 giant and the system very similar to 73~Leo  consisting of a K giant primary and an F1 secondary. Based on this information,  Glebocki \& Stawikowski (\cite{glebockistawikowski97}) adopted a mass of 2.5~M$_\odot$ for  the K giant primary, a mass ratio of 0.64, and from this got the low orbital inclination of 13$\degr$. This calculation used the fact that when the mass function is extremely low then the orbital inclination is not very sensitive on the primary mass and the mass function. The relation between the orbital inclination and mass ratio is given in Fig.~\ref{incl}, and $i$ is calculated from the following equation;\\
\begin{equation}
\sin i = K_{1}\thinspace \sqrt{1-e^2}\thinspace \frac{1}{q}\thinspace \sqrt[3]{\frac{(1+q)^2\thinspace \frac{P}{2\pi}} {Gm_1} },
\end{equation}
where $K_1$ is the amplitude of the radial velocity curve, $e$ is the eccentricity, $m_1$ is the mass of the primary, $q=m_2/m_1$ is the mass ratio in the sense that $m_1$ is the more massive star, $P$ is the orbital period and $G$ is the gravitational constant.

IT~Com is less massive, the K giant primary is about 1.4 instead of 2.5 solar masses (Table \ref{table:2}), the secondary component could not  be an F main sequence star like in 73~Leo (which would have about the same mass and would be visible in the optical spectra). From the rotational period and the radius of the primary (see Table \ref{table:1} and Table \ref{table:2}) the predicted rotational velocity is $\approx$8\,\kms . The measured $v\sin i$ is $\approx$5-6\,\kms , which implies a rotational inclination of about 40-45$\degr$. This, in turn, translates into a secondary mass of about 0.3 solar mass with the help of  Fig.~\ref{incl}, i.e., the secondary component is most likely an M dwarf. Taking into account the fairly high amplitude rotational modulation, the inclination angle can not be as low, so a coplanar orbit is probable, contrary to the result of Glebocki \& Stawikowski (\cite{glebockistawikowski97}).

\section{Masses of the secondary stars}\label{A2}

\begin{figure}[hbt]
   \includegraphics[width=5.8cm, angle=270]{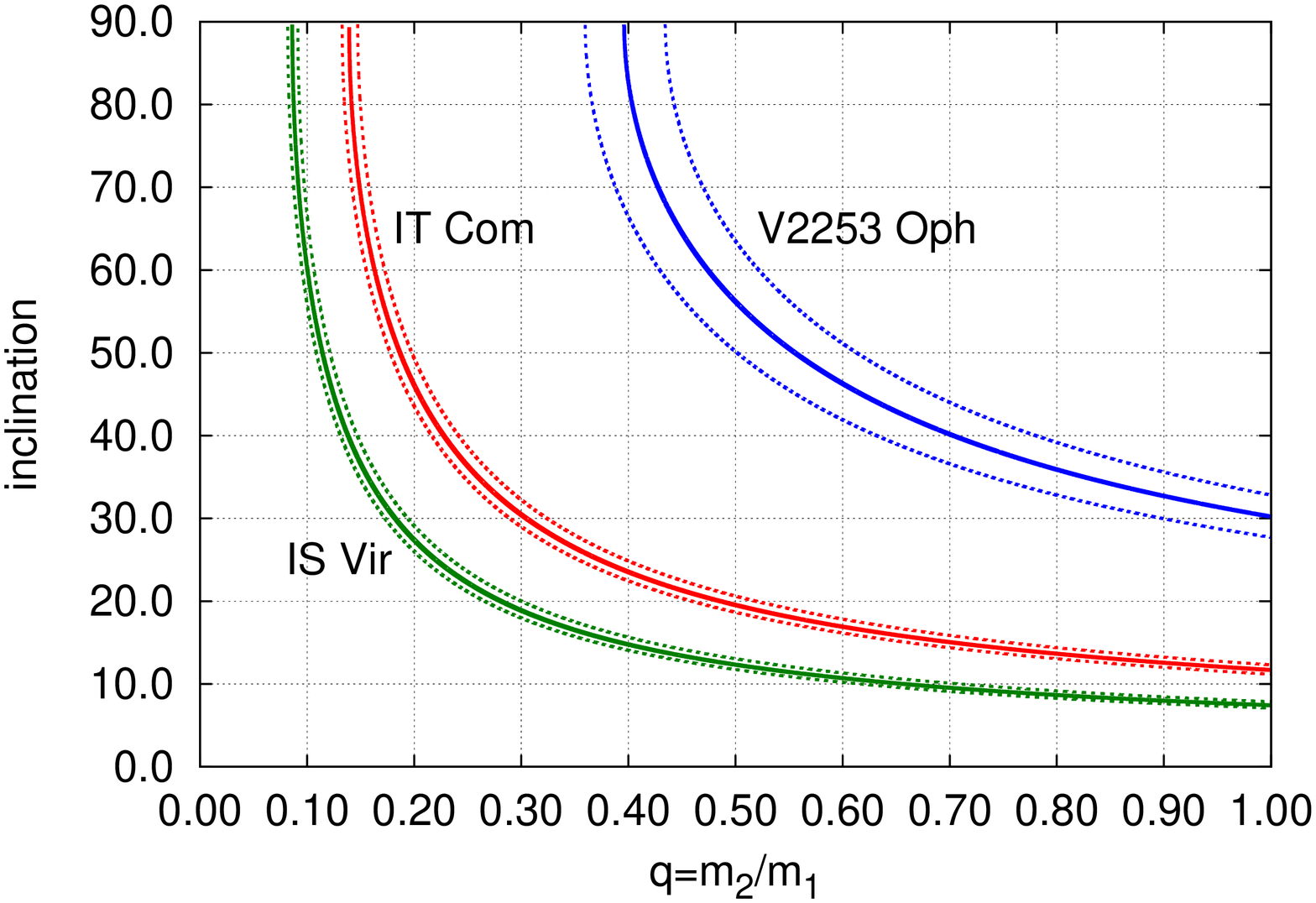}
   \caption{The relation between the orbital inclinations and mass ratios of V2253~Oph, IT~Com and IS~Vir. Solid lines mark the mean values of the primary masses, dotted lines indicate the mass ranges.}
         \label{incl1}
   \end{figure}

From the same calculations as above in appendix \ref{A1}, the possible values of the secondary masses of V2253~Oph, IT~Com and IS~Vir can be found from Fig.~\ref{incl1}. The estimated ranges of the primary masses and the orbital inclinations can be taken into account. The primary masses are as follows: V2253~Oph: 3~M$_\odot$, (range 2.4--3.8 M$_\odot$), IT~Com: 1.4~M $_\odot$, (range 1.2--1.6~M$_\odot$), IS~Vir: 1.3~M$_\odot$, (range 1.1--1.5~M$_\odot$). Eccentricities and radial velocity curve amplitudes are given in Table~\ref{table:1}.

\end{document}